\documentclass[10pt,twocolumn,letterpaper]{article}

\usepackage{cvpr}              

\usepackage{graphicx}
\usepackage{amsmath}
\usepackage{amssymb}
\usepackage{booktabs}
\usepackage{multicol}
\usepackage{multirow}
\usepackage{booktabs}
\usepackage{threeparttable}

\usepackage[pagebackref,breaklinks,colorlinks]{hyperref}

\usepackage[capitalize]{cleveref}
\crefname{section}{Sec.}{Secs.}
\Crefname{section}{Section}{Sections}
\Crefname{table}{Table}{Tables}
\crefname{table}{Tab.}{Tabs.}

\def\cvprPaperID{794} 
\def\confName{CVPR}
\def\confYear{2022}

\begin{document}

\title{Contrastive Learning for Local and Global Learning MRI Reconstruction}

\author{Qiaosi Yi,\,Jinhao Liu,\,Le Hu,\,Faming Fang\thanks{Corresponding author}\,,\,and Guixu Zhang\\
School of Computer Science and Technology, East China Normal University, Shanghai, China\\
\vspace{-10pt}
}
\maketitle

\begin{abstract}
Magnetic Resonance Imaging (MRI) is an important medical imaging modality, while it requires a long acquisition time. To reduce the acquisition time, various methods have been proposed. However, these methods failed to reconstruct images with a clear structure for two main reasons. Firstly, similar patches widely exist in MR images, while most previous deep learning-based methods ignore this property and only adopt CNN to learn local information. Secondly, the existing methods only use clear images to constrain the upper bound of the solution space, while the lower bound is not constrained, so that a better parameter of the network cannot be obtained. To address these problems, we propose a Contrastive Learning for Local and Global Learning MRI Reconstruction Network~(CLGNet). Specifically, according to the Fourier theory, each value in the Fourier domain is calculated from all the values in Spatial domain. Therefore, we propose a Spatial and Fourier Layer~(SFL) to simultaneously learn the local and global information in Spatial and Fourier domains. Moreover, compared with self-attention and transformer, the SFL has a stronger learning ability and can achieve better performance in less time. Based on the SFL, we design a Spatial and Fourier Residual block as the main component of our model. Meanwhile, to constrain the lower bound and upper bound of the solution space, we introduce contrastive learning, which can pull the result closer to the clear image and push the result further away from the undersampled image. Extensive experimental results on different datasets and acceleration rates demonstrate that the proposed CLGNet achieves new state-of-the-art results.
\end{abstract}

\section{Introduction}

Magnetic Resonance Imaging (MRI) is a widely applied medical imaging modality for disease diagnosis and treatment planning. Compared with other imaging modalities, MRI is a non-invasive and non-radiation imaging modality and can provide detailed images of the soft tissue, including structural, anatomical, and functional information. And the resolution of the MR image is affected by the number of rows/columns in k-space for MRI scanners. Therefore, to obtain a higher spatial resolution, more signals need to be collected, which will directly increase the scanning time. However, with the longer scan time, the collected signals may be degraded by motion artifacts, \ie cardiac motion, blood flow, and respiration. Moreover, the longer scan time also increases the healthcare cost to the patient and limits the availability of MR scanners~\cite{jaspan2015compressed, yang2017dagan,zhou2020dudornet, pal2021review}.

\begin{figure}[t]
  \centering
  \includegraphics[width=0.96\linewidth]{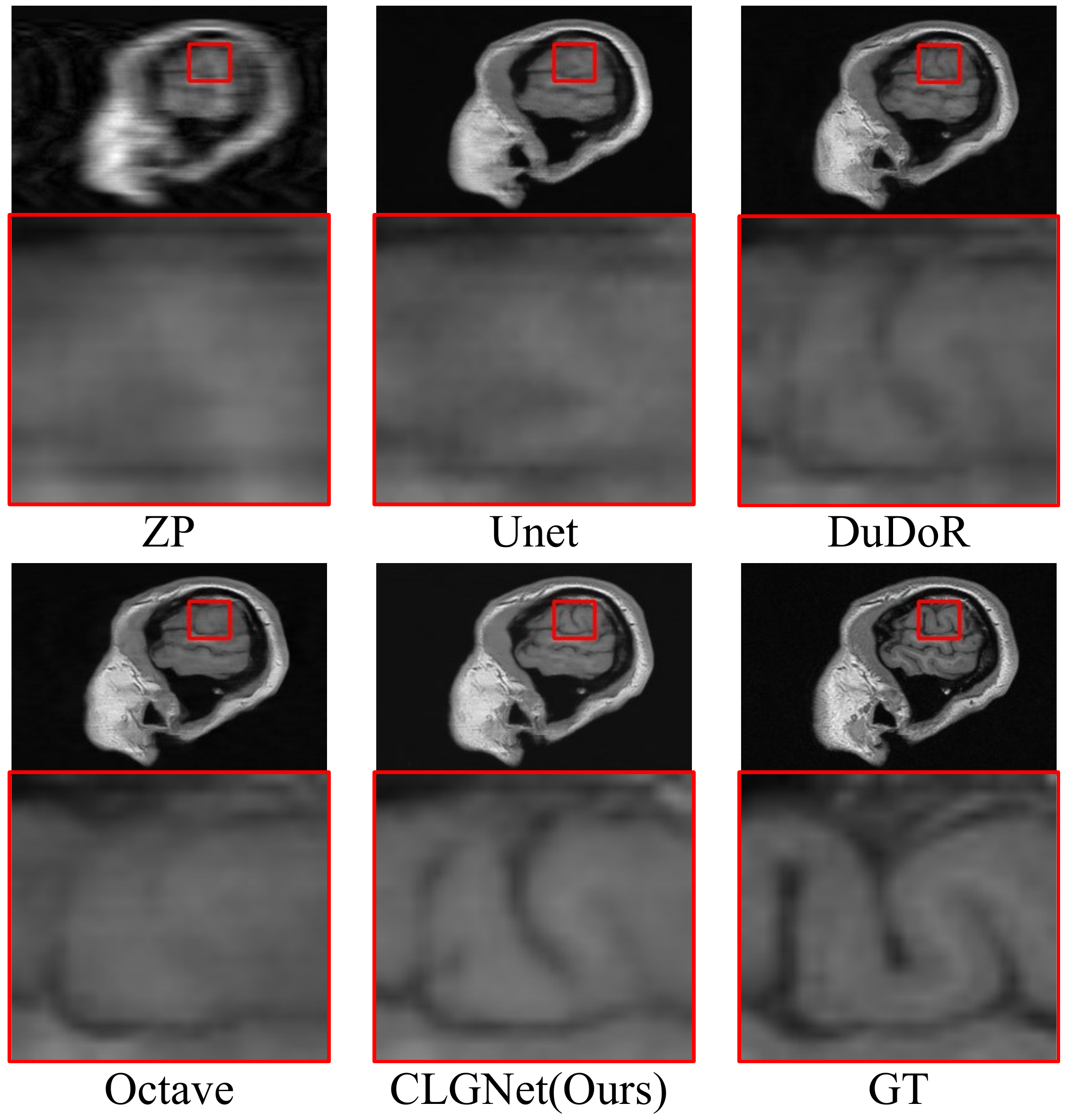}
   \caption{An example of MRI reconstruction results under $\times8$ acceleration on CC-359 Brain dataset. Obviously, CLGNet can reconstruct high-quality image with clear structure.}
   \label{fig1}
  \vspace{-10pt}
\end{figure}

To reduce the scan time, one way is to reduce the number of collected signals, that is, to reconstruct the MR image from the undersampled k-space data instead of the full k-space data. However, sampling below the Nyquist rate in the k-space domain causes aliasing artifacts and causes the undersampled image to not be completely restored to a fully-sampled image, that is, MRI reconstruction is an ill-posed problem. To better solve the ill-posed problem, many methods~\cite{liang2009accelerating,knoll2011second,jaspan2015compressed, yang2017dagan,zhou2020dudornet} have been proposed to improve the quality of MR images. These methods can be roughly divided into two categories. One is the classic Compressed Sensing-based MRI method (CS-MRI)~\cite{liang2009accelerating, qu2010combined, qu2010iterative, ravishankar2010mr, zhang2015exponential}. In CS-MRI, the sparsity of the MR image in a certain transformed domain~(\eg the wavelet domain~\cite{qu2010iterative, zhang2015exponential} and total variation~\cite{block2007undersampled, knoll2011second}) is adopted to reconstruct the aliasing-free image. However, the sparsifying transforms might be too simple to capture complex image details~\cite{yang2017dagan}. Besides, the CS-MRI usually involves iterative optimization, which is time-consuming and leads to loss of details and unwanted artifacts in high undersampling rate~\cite{yang2017dagan, zhou2020dudornet}. The other is the Deep Learning-based MRI method~(DL-MRI), which uses convolution neural network~(CNN) to directly reconstruct the aliasing-free image~\cite{yang2017dagan} (Model-free MRI) or learn the parameters of CS-MRI with unrolled architectures~\cite{ zhou2020dudornet, jun2021joint} (Model-driven MRI). Compared with CS-MRI, DL-MRI tend to have better feature representation ability, faster inference speed, end-to-end trainable paradigm, and significant performance improvement. 

Albeit DL-MRI methods have brought great performance improvements, they are difficult to reconstruct aliasing-free images with a clear structure, as shown in Fig.~\ref{fig1}. That is because: (1) The similar patches widely exist in MR images, while most methods ignore this property and only adopt CNN to learn the local information, which leads to the failure in reconstructing the high-quality aliasing-free image. Recently, some methods~\cite{wu2019self, feng2021accelerated} have adopted self-attention and transformer to learn global information, while the structure information is still missing in the recovered results. In addition, self-attention and transformer are time-consuming and memory-consuming. (2) Only using clear MR images cannot effectively constrains the solution space of the network. Given a clear image only constrains the upper bound of the solution space, while the lower bound of the solution space is not constrained. Therefore, most existing methods based on L1/L2 loss are difficult to deal with the details of images. (3) The image in Spatial domain contains more content information and the structure information of MR images is not significant. On the contrary, the image processed by wavelet transform contains more significant high-frequency information, so that the model can focus more on the structure information.

\begin{figure*}[t]
  \centering
   \includegraphics[width=0.92\linewidth]{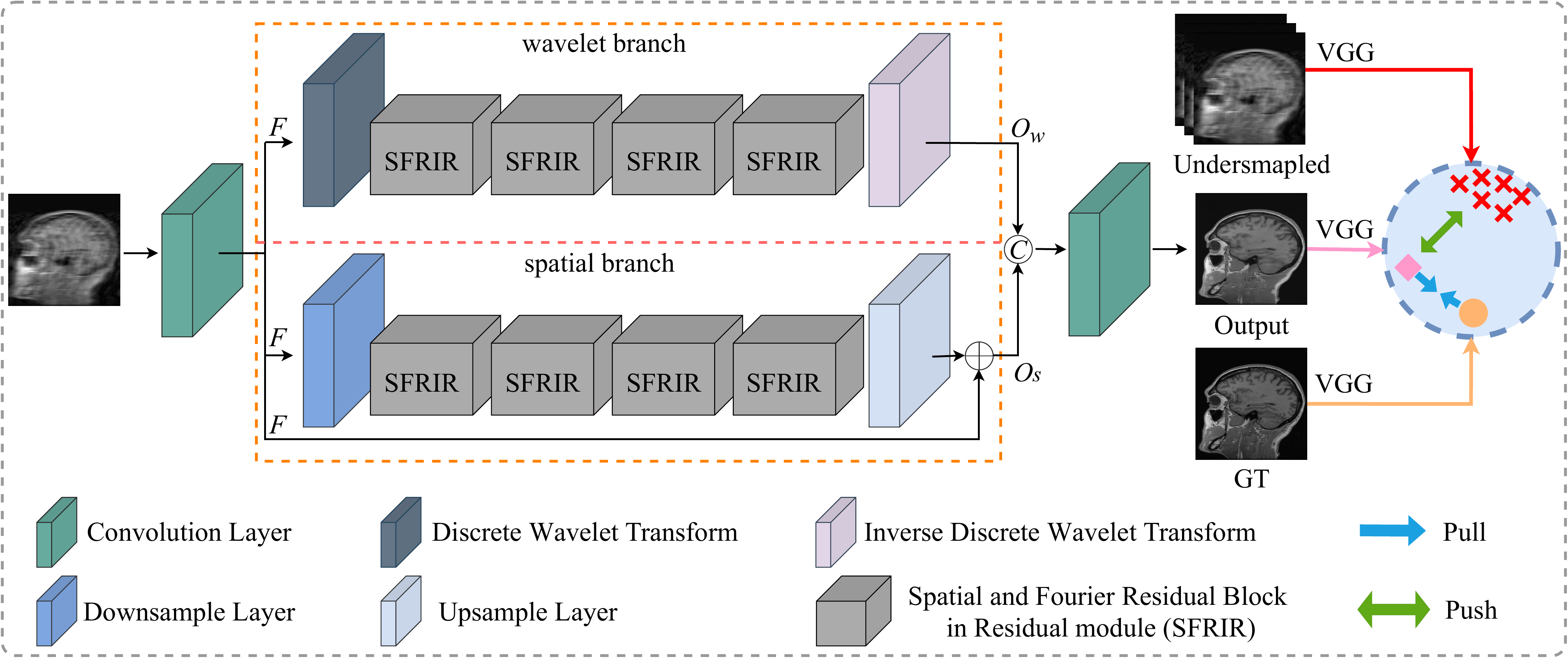}
   \caption{The architecture of our proposed Contrastive Learning  for Local and Global Learning MRI Reconstruction. }
   \label{fig:model}
   \vspace{-10pt}
\end{figure*}

To address the aforementioned issues, we propose Contrastive Learning for Local and Global Learning MRI Reconstruction network~(CLGNet), which is composed of spatial branch and wavelet branch, as illustrated in Fig.~\ref{fig:model}. Firstly, according to the Fourier theory, point-wise update in the Fourier domain globally affects all input features in the Spatial domain, which inspires us another way to learn the global information. Hence, to learn the global information, we introduce the Fast Fourier Convolution (FFC)~\cite{chi2020fast} to our model and propose the Spatial and Fourier Layer~(SFL), which adopts vanilla convolution and FFC to learn the local and global information~(see Fig.~\ref{fig:sfc}). Furthermore, based on SFL, we further propose a Spatial and Fourier Residual Block~(SFRB) and Spatial and Fourier Residual Block in Residual module (SFRIR) as main components of our model. Furthermore, compared with self-attention and transformer, the proposed SFL can learn global information more effectively in less time. Secondly, we introduce Contrastive Learning~(CL) to constrain the upper bound and lower bound of the proposed CLGNet, as shown in Fig.~\ref{fig:model}. In order to constrain the quality of the recovered image, the following two steps are required: 1) constraining the upper bound and 2) constraining the lower bound. For constraining the upper bound, we need to pull the recovered MR image closer to the clear MR image, and for constraining the lower bound, we need to push the recovered MR image farther away from the undersampled MR image. After the above two steps, the solution space will be successfully constrained and the quality of the recovered MR image will be further improved.

The main contributions of this paper are as follows:
\begin{itemize}
\item We propose a novel Contrastive Learning for Local and Global Learning MRI Reconstruction network. Extensive experimental results show that our method achieves new state-of-the-art results. In addition, the proposed CLGNet better balances the trade-off between parameters, Flops, running-time, and performance, compared to the state-of-the-art approaches.
\item We propose a Spatial and Fourier Layer~(SFL), which can simultaneously learn local and global features in Spatial domain and Fourier domain, respectively. Compared with self-attention and transformer, the proposed SFL can achieve better performance in less time.
\item To constrain the solution space of CLGNet, we introduce contrastive learning, which effectively improves the performance by constraining the distance between the result, the clear image and input image.
\end{itemize}

\section{Related Work}

\subsection{MRI Reconstruction}
To accelerate the MRI acquisition, the previous works~\cite{griswold2002generalized, huang2005k, qu2010iterative, knoll2011second} are mainly divided into parallel imaging and compressed sensing. With the renaissance of deep neural networks, deep learning have been widely used for MRI reconstruction and have achieved remarkable performance. Wang \etal~\cite{wang2016accelerating} first proposed a multi-layer CNN model for MRI reconstruction and Jin \etal~\cite{jin2017deep} adopted Unet~\cite{ronneberger2015u} to recover a clear MR image. To further improve the performance of MRI reconstruction, the cross-domain~\cite{eo2018kiki, souza2019hybrid, nitski2020cdf, zhou2020dudornet} were proposed. However, these methods separately processed the undersampled MRI in the Spatial and Frequency domains and cannot learn the local and global information, simultaneously. In contrast, to learn the global information, Wu \etal~\cite{wu2019self} proposed a self-attention network for MR imaging and Feng \etal~\cite{feng2021accelerated} designed a transformer network for MR imaging. However, the self-attention~\cite{huang2019ccnet, chen20182, wang2020linformer} and transformer~\cite{liu2021swin, sun2021rethinking} have high computation complexity and occupy a huge number of GPU memory. Moreover, the transformer~\cite{sun2021rethinking, valanarasu2021medical} is difficult to optimize and requires a large-scale training dataset. Different from them, in this work, the proposed Spatial and Fourier Layer~(SFL) can simultaneously learn the local and global information of the feature, while the required memory and the computation complexity of SFL based on the Fast Fourier Transform are similar to the convolution layer.

\subsection{Contrastive Learning}
To learn richer feature representations of data, contrastive learning~\cite{oord2018representation, he2020momentum, chen2020simple} proposed augmenting the data during training, and contrasting a model’s representations of the data by pulling samples from the same samples closer and pushing samples from different samples further away. SimCLR~\cite{chen2020simple} adopted some augmentations including spatial distortion and chromatic distortion to enrich feature representations. Considering KL divergence ignores important structural knowledge of the teacher network in knowledge distillation, Tian \etal~\cite{tian2019contrastive} adopted contrastive learning to a student so that the student can capture significantly more information in the teacher’s representation of the data. Park \etal~\cite{park2020contrastive} employed contrastive learning to improve unpaired image-to-image translation quality. Recently, Wang \etal~\cite{wang2021towards} introduced contrastive learning to improve the performance of the task of the image super-resolution. However, there are still few works to apply contrastive learning to MRI reconstruction. In this work, we aim to constrain the solution space by using the contrastive loss function, which could be achieved by pulling samples from the same samples closer and pushing samples from different samples further away.

\section{Method}

In this section, we introduce our proposed Contrastive Learning for Local and Global Learning MRI Reconstruction Network~(CLGNet) in detail. Firstly, we introduce the overall framework. Then we present the details of Spatial and Fourier Layer~(SFL), Spatial and Fourier Residual Block~(SFRB), Spatial and Fourier Residual Block in Residual module~(SFRIR), and final loss functions accordingly.

\subsection{Overview}
In MRI reconstruction, we aim to reconstruct an aliasing-free MR image $O$ from an undersampled MR image $I$. We denote our proposed method as CLGNet and its parameters as $\theta$. Hence, $O = CLGNet(I; \theta)$. To make $O$ as similar to the clear MR images $GT$ as possible, the loss function $\mathcal{L}$ is used to optimize the parameters $\theta$, as Eq.\ref{Eq:1}.
\begin{equation}
\theta^{*}=\arg \min _{\theta} \mathbb{E}_{O} \mathcal{L}\left(CLGNet\left(I; \theta\right), GT\right).
\label{Eq:1}
\end{equation}

The overall framework of our proposed CLGNet is shown in Fig.\ref{fig:model}. The proposed CLGNet can be divided into three parts: 1) the preprocessing part, 2) the spatial-wavelet learning part, and 3) the output part. The preprocessing part is mainly used to extract the feature $\mathcal{F}$ of the input images. 
\begin{equation}
\mathcal{F} = Net_{pre}(I),
\end{equation}
where $Net_{pre}$ represents the preprocessing part. 

As we analyzed above, since the previous works failed to reconstruct MR images with clear structure, we propose a spatial-wavelet learning paradigm~(SWLP) for learning content information and frequency information respectively. The SWLP is composed of spatial branch and wavelet branch. In wavelet branch, the module aims to learn the features of different frequencies and tries to recover them. Firstly, the $\mathcal{F}$ is processed to obtain different frequency features by discrete wavelet transform~(DWT). Considering the complementary information between different frequencies, we concatenate different frequencies together as the input of SFRIR. Then, the output of wavelet branch can be obtained by multiple SFRIRs and inverse discrete wavelet transform~(IDWT). 
\begin{equation}
    \left\{\begin{array}{l}
LL, LH, HL, HH = DWT(\mathcal{F}),\\
I_{s} = [LL,LH,HL,HH],\\
O_w = IDWT(SFRIR(\dotsc(SFRIR(I_{s})))),\\
\end{array}\right.
\end{equation}
where $LL,LH,HL,HH$ represent the different frequencies, $I_s$ is the input of SFRIR, $[\dotsc]$ represents the concatenation operator, and $O_w$ is the output of wavelet branch. In wavelet branch, we set the number of SFRIR to 4. In spatial branch, the module aims to learn the content of images. Moreover, the architecture of spatial branch is similar to the wavelet branch. The difference is that we adopt skip-connection in spatial branch to prevent information loss and use Downsample layer and Upsample layer to replace DWT and IDWT. 

In the output part, the output results of spatial branch and wavelet branch are concatenated together and processed by the convolution layer to obtain the final output result $O$. 
\begin{equation}
O = Net_{out}([O_w,O_s]),
\end{equation}
where $Net_{out}$ represents the output part, $O_w$ is the output of wavelet branch, and $O_s$ is the output of spatial branch. Since skip-connection is used in spatial branch, the input of the output part already includes $\mathcal{F}$. Furthermore, since $\mathcal{F}$ is extracted from the input images, $\mathcal{F}$ does not have obvious structural information. Hence, in order not to interfere with learning in wavelet branch, we do not use skip-connection in wavelet branch.

\begin{figure}[t]
  \centering
   \includegraphics[width=1\linewidth]{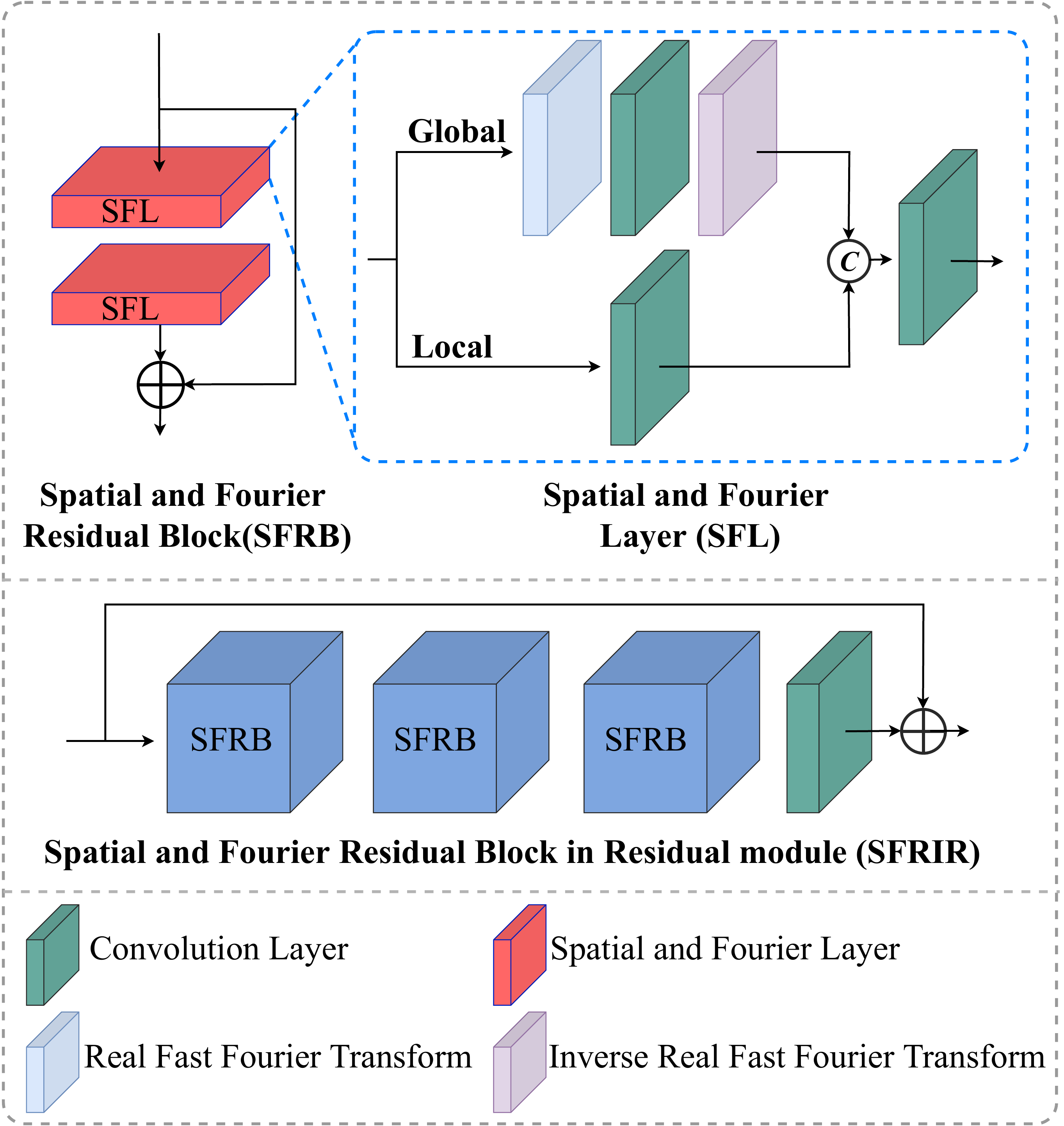}
   \caption{The architecture of our proposed Spatial and Fourier Layer, Spatial and Fourier Residual block, and Spatial and Fourier Residual Block in Residual module.}
   \label{fig:sfc}
   \vspace{-10pt}
\end{figure}
\subsection{Spatial and Fourier Layer}
As we know, similar patches widely exist in MR images. However, only using dual-domain learning paradigm cannot effectively learn global information. Therefore, we further propose Spatial and Fourier layer~(SFL) as the basic layer to effectively learn local and global information.

\subsubsection{Fourier transform}

\begin{table*}[htbp]
	\centering
	\setlength{\tabcolsep}{1.9mm}
		\centering
			\begin{tabular}{lccccccccccccccc}
				\toprule
				\multirow{2}{*}{Methods} &
				\multicolumn{3}{c}{CC-359 Brain (4$\times$)} & \multicolumn{3}{c}{CC-359 Brain (8$\times$)} & \multicolumn{3}{c}{FastMRI (x4)}&\multicolumn{3}{c}{FastMRI (8$\times$)}\cr
				\cmidrule(lr){2-4} \cmidrule(lr){5-7} \cmidrule(lr){8-10} \cmidrule(lr){11-13} 
				    & NMSE & PSNR & SSIM & NMSE & PSNR & SSIM & NMSE & PSNR & SSIM & NMSE & PSNR & SSIM\cr
				\midrule
				ZP    &  0.0483&26.34 & 0.7510&  0.1066& 22.85 & 0.6424 &0.0707  &29.93  &0.6583  & 0.1105 &27.42 &0.5546 \cr
				Unet   & 0.0213&29.90&0.8511&0.0456&26.61&0.7705 & 0.0449&32.00&0.7178&0.0591&30.24&0.6478\cr
				DCCNN & 0.0082&34.32&0.9163&0.0274&28.91&0.818 & 0.0441&32.32&0.7228&0.0586&30.33&0.6488 \cr
				ISTA  & 0.0173&30.71&0.8654&0.0395&27.08&0.7778 & 0.0455&31.82&0.7146&0.0626&29.80&0.6404\cr
				CDDN  & 0.0114&32.85&0.8972&0.0408&27.25&0.7852 & 0.0563&31.30&0.6977&0.0715&29.50&0.6217\cr
				PD  & \underline{0.0071}&\underline{35.02}&\underline{0.9226}&0.0284&28.79&0.8199 & \underline{0.0437}&32.17&0.7223&0.0581&30.28&0.6506\cr
				DuDoR & 0.0076&34.64&0.9205&\underline{0.0244}&\underline{29.40}&\underline{0.8309} & ----- & ----- & ----- & ----- & ----- & -----\cr
				OctConv &0.0079&34.53&0.9178&0.0253&29.19&0.8249 & 0.0438&\underline{32.40}&\underline{0.7259}&\underline{0.0565}&\underline{30.60}&\underline{0.6557}\cr
				CLGNet & \textbf{0.0063} & \textbf{35.63} & \textbf{0.9294} & \textbf{0.0232} & \textbf{29.80} & \textbf{0.8375} & \textbf{0.0421} & \textbf{32.54} & \textbf{0.7313} &\textbf{0.0537}&\textbf{30.88}&\textbf{0.6640}\cr
				\bottomrule
			\end{tabular}
			\caption{Quantitative experiments evaluated on different datasets and acceleration rates. The best and the second best results have been boldfaced and underlined. ZP is the zero-padding and DuDoR is the DuDoRNet withour T1 prior, since CC-359 Brain and FastMRI datasets do not include the corresponding T1 MR images.}
			\label{Tab:compare}
			\vspace{-10pt}
\end{table*}

According to Fourier theory, for a 2D matrix $f\in \mathbb{R}^{m\times n}$, its Fourier transform can be expressed as:
\begin{equation}
F(u, v)=\sum_{x=0}^{m-1} \sum_{y=0}^{n-1} f(x, y) e^{-j 2 \pi\left(\frac{\mathrm{ux}}{\mathrm{m}}+\frac{v y}{n}\right)},
\label{eqfourier}
\end{equation}
where $F$ is the data in Fourier domain, $u$ and $v$ is the index of $F$, and $x$ and $y$ is the index of $f$. According to Eq.\ref{eqfourier}, we can observe that each value in Fourier domain depends on all values in Spatial domain. Hence, the size of the receptive field of a convolution layer in Fourier domain is the size of the entire image, so that the global information can be learned in Fourier domain. In addition, since each value in MR images contains a real part and an imaginary part, the previous methods~\cite{cheng2019model,zhou2020dudornet} concatenate the real part and the imaginary part as the input. However, the features extracted by convolution are learned from the real part and the imaginary part together, which is not conducive to using Fourier transform on the features. Therefore, similar to some works~\cite{jin2017deep, feng2021task}, we calculate the absolute value of MR images as the input and use real Fast Fourier transform~(\textit{torch.fft.rfft}) to replace Fast Fourier transform~\cite{chi2020fast}. 

\subsubsection{Spatial and Fourier Layer}
As shown in Fig.~\ref{fig:sfc}, the SFL contains local branch and global branch. In local branch, the vanilla convolution is adopted to learning the local information.
\begin{equation}
F_L = Conv(F_s),
\end{equation}
where $F_s$ is the input of SFL, $Conv$ represents the convolution layer, and $F_L$ is the output of local branch. In global branch, the real Fast Fourier transform is used to convert $F_s$ to $F_t$, which is the data in Fourier domain. According to Eq.\ref{eqfourier}, we can know that each value in $F_t$ depends on all values in $F_s$ and the global information of $F_s$ can be learned by using convolution layer on $F_t$, thereby the global branch can be expressed as: 
\begin{equation}
F_G = IFFT(Conv(FFT(F_s))),
\label{eqg}
\end{equation}
where $FFT$ is the real Fast Fourier transform, $IFFT$ is the inverse real Fast Fourier transform, $F_G$ represents the output of the global branch. After obtaining $F_L$ and $F_G$,  the output $O_s$ of SFL can be obtained:
\begin{equation}
O_s = Conv([F_L, F_G]).
\label{eqg}
\end{equation}
Furthermore, inspired by Residual Block~(RB)~\cite{he2016deep} and Residual in Residual~(RIR)~\cite{zhang2018image}, we propose the Spatial and Fourier Residual Block~(SFRB) and the Spatial and Fourier Residual Block in Residual module~(SFRIR) as the main components of the proposed CLGNet, as illustrated in Fig.~\ref{fig:sfc}. In SFRB, we use two SFL to learn the information and adopt skip-connection to prevent information loss. In addition, in SFRIR, we use SFRB to replace the attention residual block in RIR and set the number of SFRB to 3.

\begin{figure*}[t]
  \centering
  \includegraphics[width=1\linewidth]{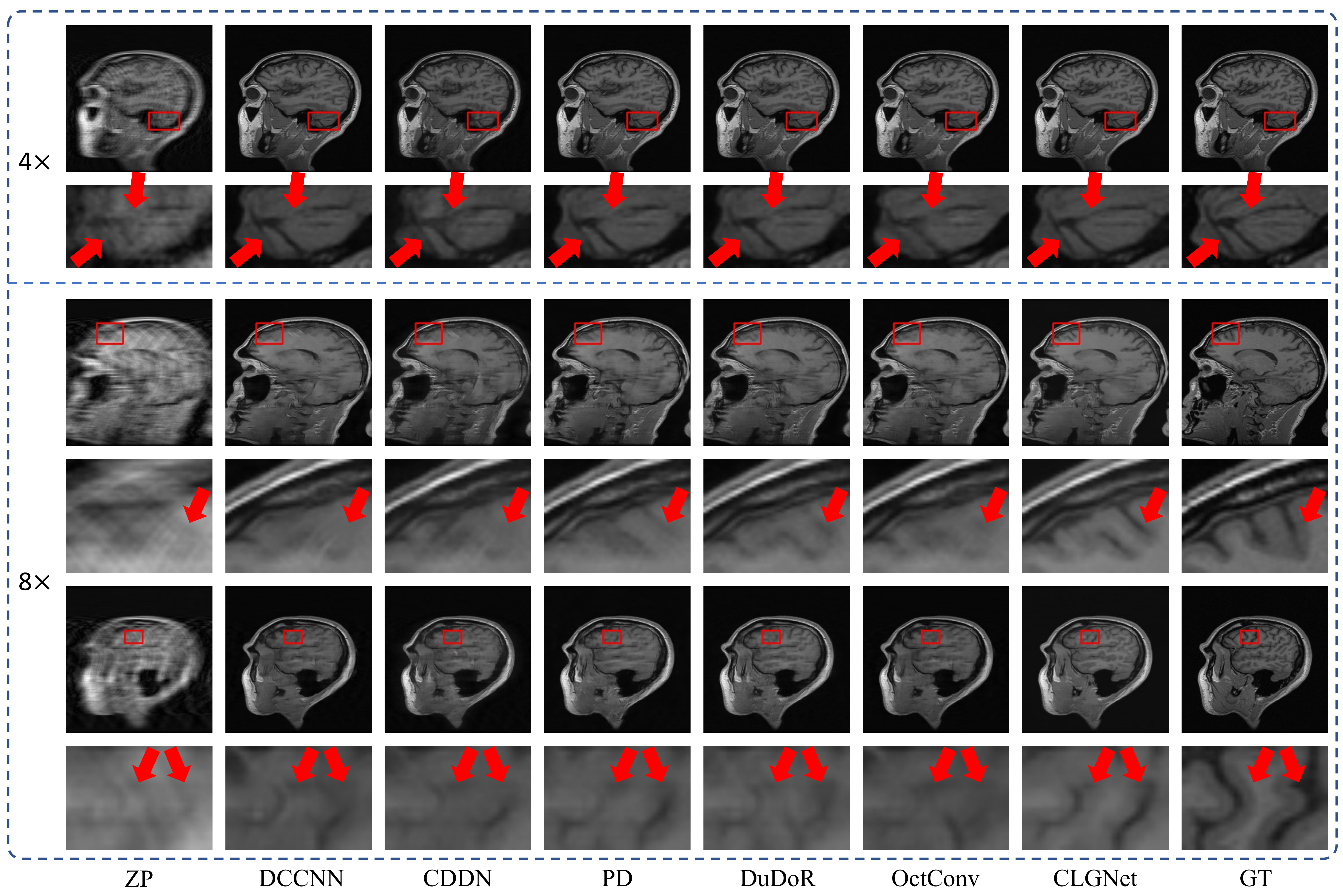}
   \caption{Reconstruction results under $4\times$ and $8\times$ acceleration on CC-359 Brain dataset. It is obvious that the proposed CLGNet can reconstruct high-quality aliasing-free MR images with clear structure.}
   \label{fig:brain}
   \vspace{-15pt}
\end{figure*}

\subsection{Loss function}

L1/L2 loss are commonly used to train the network, while they only constrain the upper bound of the solution space and the lower bound is neglected, thus they are difficult to reconstruct a high-quality MR image. To constrain the upper bound and lower bound of the solution space, we introduce contrastive learning, which can pull anchors close to positive samples while push away negative samples. For contrastive learning, how to select positive and negative samples and how to find the potential feature space are very important. In our work, we denote the output of CLGNet and the clear image as the anchor and the positive sample, respectively. For negative samples, following \cite{wang2021towards}, we sample $K$ undersampled MR images from the same mini-batch as negative samples. For the latent feature space, VGG19~\cite{simonyan2014very} is adopted to extract the feature of anchor~($O$), positive~($GT$), and negative~($O_{Neg}$) samples. Hence, the contrastive loss can be expressed as: 
\begin{equation}
\small
L_{CL}=\sum_{i}^{N} \sum_{j}^{M} \lambda_{j} \frac{||(\phi_{j}(O^{(i)}), \phi_{j}(GT^{(i)})||_1}{\sum_{k}^{K} ||(\phi_{j}(O^{(i)}), \phi_{j}(O_{N e g}^{(k)})||_1},
\label{cl}
\end{equation}
where $\phi_{j}$ represents the extracted feature of $j$-th layer of VGG19 and $\lambda_j$ is the balancing weight for each layer. By minimizing Eq.~\ref{cl}, the result will be close to the clear image and far away from the input image. In addition, we also use L1 loss to constrain the quality of the reconstructed image. Hence, the total loss function of the proposed CLGNet combines the L1 loss and $\mathcal{L}_{CL}$:
\begin{equation}
\mathcal{L} = ||GT-O||_1 + \alpha \mathcal{L}_{CL},
\label{eqsegloss}
\end{equation}
where $\mathcal{L}$ is the total loss, $GT$ is the clear MR images, $O$ is the reconstructed MR images, and $\alpha$ is a hyperparameter.

\section{Experiments}

\subsection{Datasets}

\textbf{Reconstructed datasets:} In our experiment, we adopt two Raw MR image datasets to train our proposed CLGNet and the other methods: (1) Calgary-Campinas-359 Brain Reconstructed data~(CC-359 Brain)~\cite{warfield2004simultaneous} provides 45 fully-sampled T1-weighted Brain MR subjects acquired on a clinical MR scanner (Discovery MR750; General Electric (GE) Healthcare, Waukesha, WI) and the matrix size is $256 \times 256$. In addition, the CC-359 dataset includes the training dataset~(25 subjects), the validation dataset~(10 subjects), and the test dataset~(10 subjects). (2) FastMRI~\cite{zbontar2018fastmri} is the largest open-access raw MR image dataset. In our work, we select the single-coil Knee data to train and test the proposed CLGNet and the other compared methods. The single-coil Knee data include 973 training subjects and 199 validation subjects, while the test subjects are not provided. Hence, we evaluate the performance of our proposed CLGNet and the other methods on the validation subjects. 

\textbf{Implementation Details:} We set $K$ as 6 and set $\alpha$ as 0.05. The learning rate is set to $1\times 10^{-4}$. We use Adam optimizer with batch size of 10 for training on four NVIDIA Titan Xp GPU. Following FastMRI~\cite{zbontar2018fastmri}, we use random cartesian undersampling to obtain the undersampled MR images. 

\begin{figure*}[t]
  \centering
  \includegraphics[width=1\linewidth]{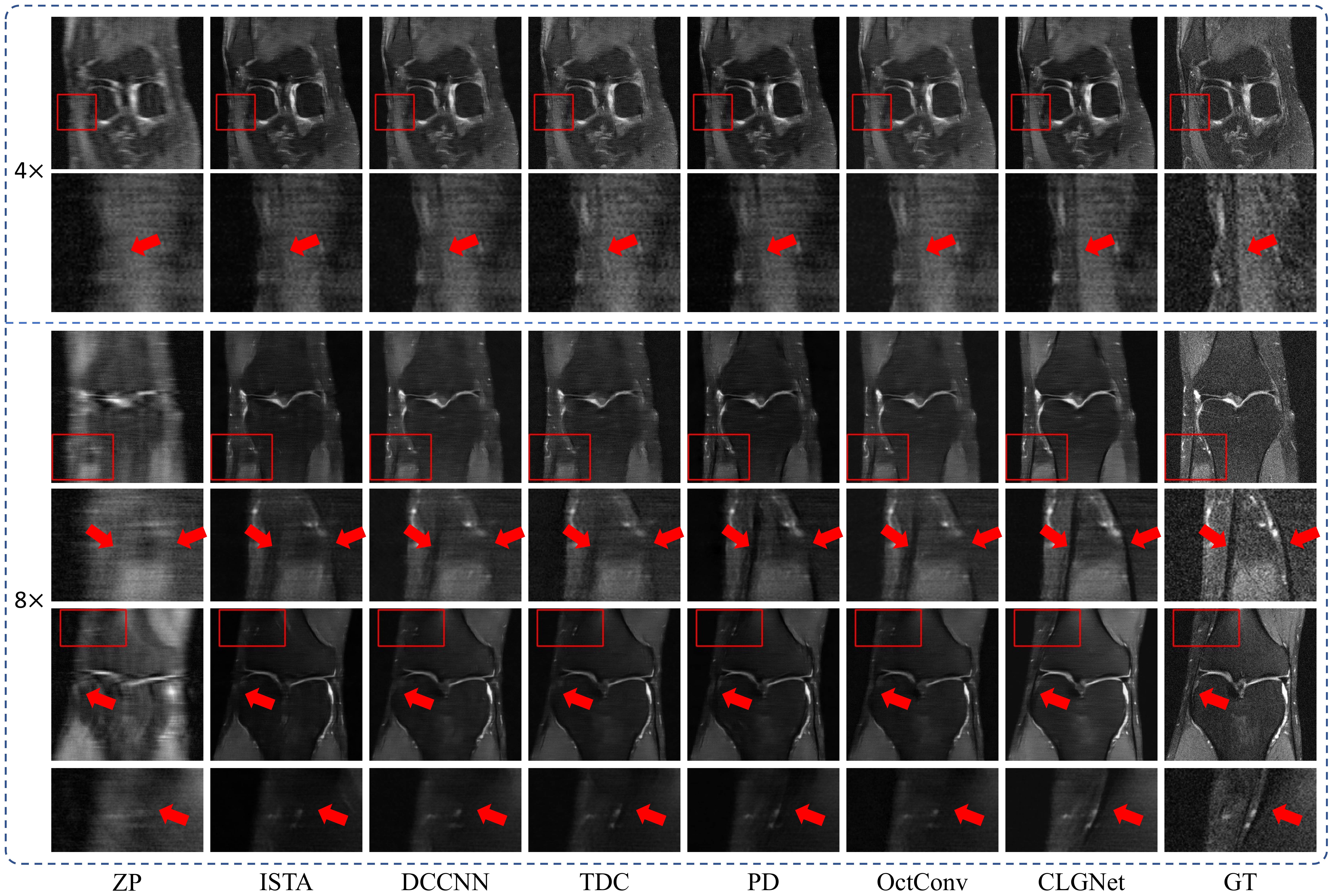}
   \caption{Reconstruction results under $4\times$ and $8\times$ acceleration on FastMRI dataset. It is obvious that the proposed CLGNet can reconstruct high-quality aliasing-free MR images with clear structure.}
   \label{fig:fastmri}
  \vspace{-15pt}
\end{figure*}
\begin{table}[t]
	\centering
	\setlength{\tabcolsep}{0.8mm}
		\centering
			\begin{tabular}{lccccc}
				\toprule
				\multirow{2}{*}{Methods}& \multirow{2}{*}{Param.} & \multirow{2}{*}{Flops} & \multirow{2}{*}{Time} &  \multicolumn{2}{c}{Brain (4$\times$)}\cr
				\cmidrule(lr){5-6}
				       & (M) & (GMac)  & (s) & PSNR & SSIM \cr
				\midrule
				Unet    & 7.76 & 54.82 & 0.0131 &29.90&0.8511\cr
				DCCNN    & 0.57 & 148.65 & 0.0714 &34.32&0.9163\cr
				ISTA & 0.38 & 148.57 & 0.0579 &30.71&0.8654\cr
				CDDN & 0.12 & 32.05 & 0.5380 &32.85&0.8972\cr
				PD  & 0.41 & 107.72 & 0.0431 &\underline{35.02}&\underline{0.9226}\cr
				DuDoR & 0.65 & 851.53 & 35.5164 &34.64&0.9205\cr
				OctConv & 0.57 & 471.07 & 0.5397 &34.53&0.9178\cr
				CLGNet(Ours) & 0.96 & 68.91 & 0.1129 &\textbf{35.63} & \textbf{0.9294}\cr
				\bottomrule
			\end{tabular}
			\caption{Comparison results of parameters, Flops and running-time on CC-359 Brain dataset. The running-time is an average time,
which is calculated on 100 images of $512 \times 512$ size.}
			\label{Tab:param}
			\vspace{-20pt}
\end{table}

\begin{table}[t]
	\centering
	\setlength{\tabcolsep}{1.5mm}
		\centering
			\begin{tabular}{lccccc}
				\toprule
				\multirow{2}{*}{Methods}& \multirow{2}{*}{Wavelet} & \multirow{2}{*}{SFL} & \multirow{2}{*}{$\mathcal{L}_{CL}$} &  \multicolumn{2}{c}{Brain (4$\times$)}\cr
				\cmidrule(lr){5-6}
				       &  &  & & PSNR & SSIM \cr
				\midrule
				w/o wavelet    &  & $\checkmark$ & $\checkmark$ &35.51&0.9261\cr
				w/o SFL    & $\checkmark$ &  & $\checkmark$ &31.65&0.8835\cr
				w/o $\mathcal{L}_{CL}$ & $\checkmark$ & $\checkmark$ & & 35.50 &0.9284\cr
				w/o negative & $\checkmark$ & $\checkmark$ & $\checkmark$ & 35.57 & 0.9286 \cr
				CLGNet(Ours) & $\checkmark$ & $\checkmark$ & $\checkmark$ & 35.63 & 0.9294\cr
				\bottomrule
			\end{tabular}
			\caption{Ablation study on different settings of CLGNet under $4\times$ acceleration on CC-359 Brain dataset.}
			\label{Tab:com}
			\vspace{-15pt}
\end{table}
\subsection{Comparison with the State-of-the-Arts}

\begin{table*}[t]
	\centering
		\centering
			\begin{tabular}{lccccccccc}
				\toprule
				\multirow{2}{*}{Methods}& \multirow{2}{*}{Self-attention} & \multirow{2}{*}{Transformer} & \multirow{2}{*}{SFL} & \multirow{2}{*}{Param.} & \multirow{2}{*}{Flops} & \multirow{2}{*}{Time} &  \multicolumn{3}{c}{CC-359 Brain (4$\times$)}\cr
				\cmidrule(lr){8-10}
				       &  &  & &(M)&(GMac)&(s)& NMSE &PSNR & SSIM  \cr
				\midrule
				CLGNet$_{non}$  &&&&0.81&65.1&0.0852 & 0.0145 &31.65&0.8835\cr
				CLGNet$_{SA}$    & $\checkmark$ &  &&0.88&66.73&0.6220& 0.0143 &31.78& 0.8841\cr
				CLGNet$_{Tran}$    &  & $\checkmark$ & &0.95&82.56&0.2993 &0.0140&32.01&0.8858\cr
				CLGNet(Ours) &  & & $\checkmark$ &0.96&68.91&0.1129 &0.0063 & 35.63 & 0.9294 \cr
				\bottomrule  
			\end{tabular}
			\caption{Explore the influence of the proposed SFL under $4\times$ acceleration on CC-359 Brain dataset.}
			\label{Tab:sfl}
			\vspace{-19pt}
\end{table*}

We compare our method with the state-of-the-art MRI reconstruction methods, including Unet\cite{jin2017deep}, DCCNN~\cite{schlemper2017deep}, ISTA~\cite{zhang2018ista}, CDDN~\cite{zheng2019cddnet}, PD~\cite{cheng2019model}, DuDoR~\cite{zhou2020dudornet}, and OctConv~\cite{feng2021dual}. Quantitative results are shown in Table~\ref{Tab:compare}. We evaluate our proposed CLGNet under $4\times$ and $8\times$ acceleration on CC-359 Brain dataset and FastMRI dataset, respectively. It can be seen that our proposed CLGNet achieves remarkable improvements over these state-of-the-art methods on different datasets and acceleration rates. This substantiates the flexibility and generality of our proposed method on different organs and different acceleration rates. In addition, Table.~\ref{Tab:param} shows the comparison results of parameters, Flops, and running-time. The running-time is an average time, which is calculated on 100 images of $512 \times 512$ size. Compared with DudoR~\cite{zhou2020dudornet} and OctConv~\cite{feng2021dual}, although our proposed CLGNet increases parameters, the running-time and Flops are significantly lower than these two methods. Compared with CDDN~\cite{zheng2019cddnet}, although our proposed CLGNet increases parameters and Flops, the proposed CLGNet achieves a better performance in less time. In addition, compared with ISTA~\cite{zhang2018ista} and PD~\cite{cheng2019model}, the CLGNet achieves remarkable improvements in fewer Flops. Therefore, the CLGNet better balances the trade-off between parameters, Flops, running-time, and performance than these top-performing methods.

Fig.~\ref{fig:brain} and Fig.~\ref{fig:fastmri} illustrate the MRI reconstruction performance of all competing methods on different datasets and different acceleration rates. As shown, the reconstruction result of our proposed is better than that of other methods in finely recovering the structure of MR images. For other comparison methods, they tend to blur the image textures.

\subsection{Ablation Study}

\textbf{Effectiveness of Basic Components:} We investigate the effects of each key component of our proposed CLGNet. The results are shown in Table.~\ref{Tab:com}. Firstly, for wavelet branch, we remove the module of wavelet branch, while for a fair comparison, we adopt the module of spatial branch to replace the module of wavelet branch, thereby keeping the network parameters unchanged. From Table.~\ref{Tab:com}, compared with CLGNet without Wavelet branch, our proposed CLGNet achieves significant performance improvements, which indicates that reconstruction in wavelet branch can help to improve the quality of the reconstructed image. Secondly, we only remove FFT and IFFT in SFL to evaluate the performance of SFL for a fair comparison. From Table.~\ref{Tab:com}, CLGNet with SFL achieves huge performance improvements~(about 4dB in PSNR) over CLGNet without SFL, which demonstrates that SFL has a strong learning ability and effectively learn local and global information. Thirdly, to constrain the solution space, we introduce contrast learning. In order to demonstrate the effectiveness of contrastive learning, we conduct an ablation experiment by removing the contrastive learning loss $\mathcal{L}_{CL}$, as shown in Table.~\ref{Tab:com}. CLGNet with $\mathcal{L}_{CL}$ outperforms CLGNet without $\mathcal{L}_{CL}$ in NMSE, PSNR, and SSIM, which demonstrates that a better solution can be obtained under the constraints of contrastive learning. In addition, to further verify the effectiveness of constraining the lower bound of solution space, we remove the negative sample and only use the positive sample. We denote the ablation experiment as $w/o negative$ in Table.~\ref{Tab:com}. Compared with CLGNet without negative, the proposed CLGNet achieves a higher performance, which further indicates that constraining the lower bound is conducive to finding a better solution in the solution space.

\begin{figure}[t]
  \centering
  \includegraphics[width=0.8\linewidth]{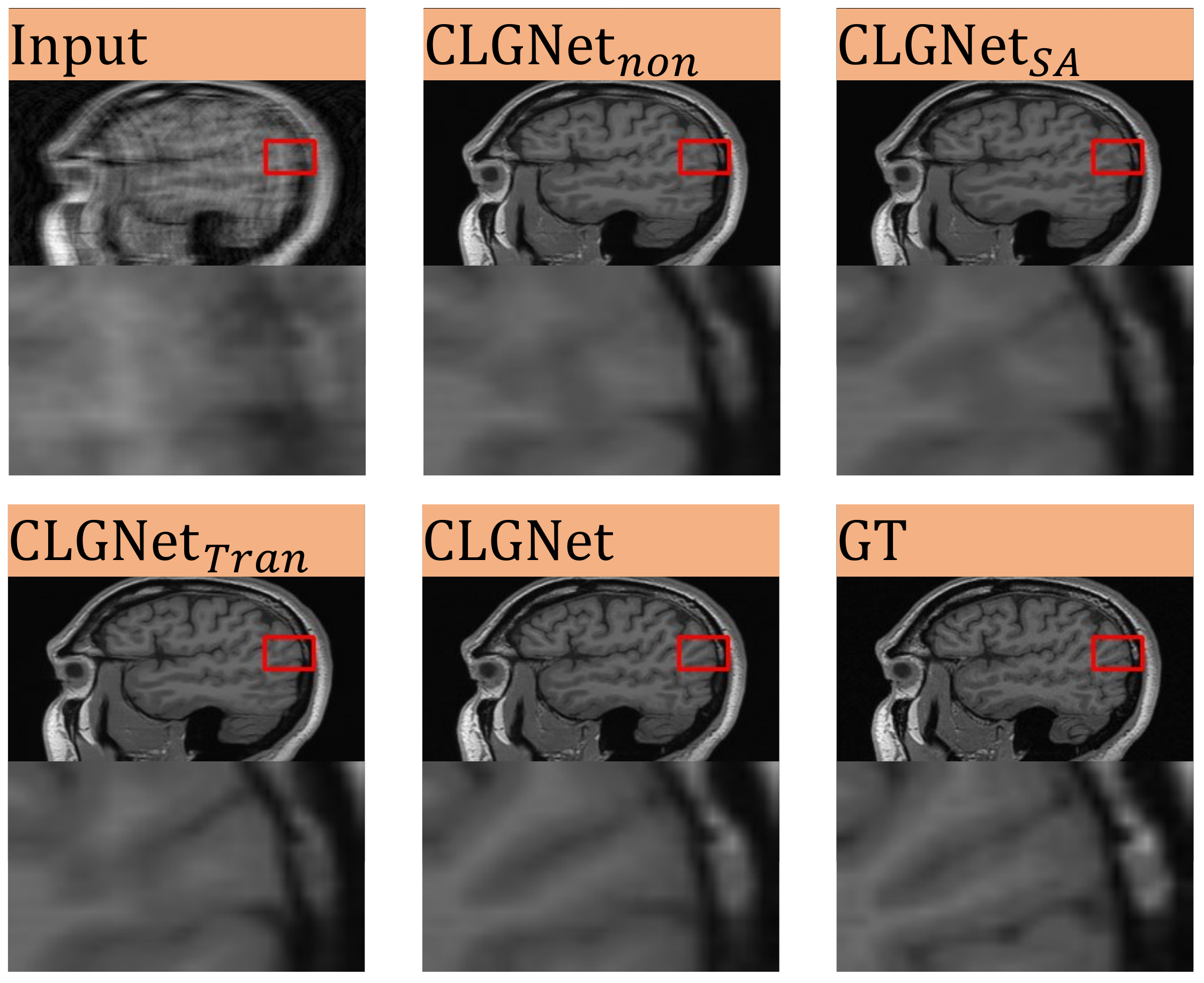}
  \vspace{-10pt}
   \caption{Comparison results of different methods of learning global information. }
   \label{fig:sflSelf}
  \vspace{-20pt}
\end{figure}

\textbf{Effectiveness of Spatial and Fourier Layer:} To further demonstrate the strong learning ability of SFL, we introduce self-attention and transformer to replace the SFL. The results are shown in Table.~\ref{Tab:sfl}.  The running-time is an average time, which is calculated on 100 images of $512\times 512$. The CLGNet$_{non}$ is the CLGNet without SFL, self-attention, and Transformer, which only learn the local information. CLGNet$_{SA}$ is the CLGNet with self-attention and CLGNet$_{Tran}$ is the CLGNet with transformer. For CLGNet$_{SA}$, we only add the self-attention in the end of SFRIR, since self-attention is time-consuming and memory-consuming. In addition, for CLGNet$_{Tran}$, we use the  Residual Swin Transformer Block proposed~(RSTB) in SwinIR~\cite{liang2021swinir} to replace the SFRIR. For a fair comparison, we try to keep the parameters of all models consistent. From Table.~\ref{Tab:sfl}, our proposed CLGNet, CLGNet$_{SA}$, and CLGNet$_{Tran}$ outperform CLGNet$_{non}$ in NMSE, PSNR, and SSIM, which shows that learning global information is essential for MRI reconstruction. Moreover, compared with CLGNet$_{SA}$ and CLGNet$_{Tran}$, our proposed CLGNet achieves significant performance improvements and the running-times of our proposed CLGNet is also less than CLGNet$_{SA}$ and CLGNet$_{Tran}$, which indicates that SFL can learn global information more effectively than self-attention and Transformer in less time. Meanwhile, compared CLGNet$_{non}$, the running-time and Flops of the proposed CLGNet only increase a little, which indicates that SFL based on FFT only slightly increases calculation and running-time compared to the convolution layer. Hence, from Table.~\ref{Tab:sfl}, it is observed that SFL has a strong learning ability and only adds a little running-time and calculation. In addition, Fig.~\ref{fig:sflSelf} illustrates the results of different methods. It is obvious that the proposed CLGNet can reconstruct high-quality MR images than other CLGNet$_{SA}$ and CLGNet$_{Tran}$, which indicates that SFL can learn global information more effectively.

\section{Conclusion}
In this paper, we have proposed a Contrastive Learning for Local and Global Learning MRI Reconstruction Network~(CLGNet), which is composed of spatial branch and wavelet branch. Specifically, we propose Spatial and Fourier Layer (SFL) to simultaneously learn the local and global information in Spatial and Fourier domains. Based on the SFL, we design Spatial and Fourier Residual block as the main components of our model. Meanwhile, in order to constrain the upper and lower bounds of the solution space, we introduce contrastive learning, which can pull the result closer to the clear image and push the result farther away from the undersampled image. Extensive experimental results demonstrate that the proposed SFL has a stronger learning ability and can achieve better performance in less time, compared with self-attention and transformer. In addition, the ablation study also demonstrates the effectiveness of contrastive learning and the wavelet branch. Extensive experiments quantitatively and qualitatively demonstrate the advantages of our proposed CLGNet over other top-performing methods.

\newpage
{\small
\bibliographystyle{ieee_fullname}
\bibliography{egbib}
}

\end{document}